\documentstyle[12pt]{article}
\newcommand{\gsim}{\mbox{\raisebox{-.3em}{$\stackrel{>}{\sim}$}}}
\newcommand{\lsim}{\mbox{\raisebox{-.3em}{$\stackrel{<}{\sim}$}}}

\renewcommand{\cite}[1]{\ref{#1}}
\newcommand{\half}{\frac{1}{2}}
\newcommand{\psibar}{\overline{\psi}}\newcommand{\psh}{p\hspace{-.4em}/}

\begin{document}
\baselineskip=0.6cm

\begin{titlepage}
\vspace*{4cm}
\begin{center}
{\Large\bf Induced Violation of Weak Equivalence\vspace{0.4cm}\\
 Principle in the Brans-Dicke Theory}\vspace{1cm}\\
Yasunori Fujii\footnotemark\footnotetext{E-mail:
ysfujii@tansei.cc.u-tokyo.ac.jp}\\
Nihon Fukushi University\\
Okuda, Chita-gun, Aichi, 470-32\ Japan\\
and\\
Institute for Cosmic Ray Research \\
University of Tokyo\\
Tanashi, Tokyo, 188 Japan \vspace{2cm}\\
\end{center}
\begin{abstract}
A quantum correction to the Brans-Dicke theory due to interactions among
matter fields is calculated, resulting in violation of WEP,
hence giving a constraint on the parameter $\omega$ far more stringent than
accepted so far.
The tentative estimate gives the lower bounds $\gsim 10^{6}$ and $\gsim 10^{8}$
 for the assumed force-range $\gsim 1$m and $\gsim 1$AU, respectively.
\end{abstract}
\end{titlepage}

Many aspects of the Brans-Dicke (BD) theory [\cite{jbd}], as a classical
theory, have been the focus of extensive studies.
Little attention seems to have been paid, however, to quantum effects due to
the interaction among matter fields.
This might be serious, because it results in violation of Weak Equivalence
Principle (WEP)\footnote{WEP, sometimes called composition-independence,
implies that any object at a given spacetime point falls with a common
acceleration independent of its mass and composition.
See, for example, C. Will, Phys. Rep. {\bf 113}, 346 \ (1984).} to the extent
that the bound on the parameter $\omega$ will be
made much more stringent than accepted so far, as will be demonstrated on the
basis of a simple model.

Assume the Lagrangian:
\begin{equation}
{\cal L}=\sqrt{-g}\left[ \frac{1}{8\omega}\phi^{2}R
-\frac{1}{2}g^{\mu\nu}\partial_{\mu}\phi\partial_{\nu}\phi +L_{\rm matter}
\right],
\label{cp_1}\end{equation}
where our $\phi$, related to the original symbol by
$\varphi_{BD}=\phi^{2}/8\omega$, fits the conventional notation in the
standard quantum field theory.
The solar system experiments have given a widely accepted lower bound
$\omega\gsim 500$.

The nonminimal couplings  of the scalar field essentially of the same nature
are known to be rather common in many of the theoretical models of unification,
 thus making this classic model particularly relevant.

The assumed masslessness of the scalar field is, however, vulnerable against
quantum effects; there is no principle that prevents a scalar field from
acquring a nonzero mass.
It is even likely that the acquired mass-squared is $\mu^{2}\sim Gm^{4}$,
giving a force-range $\lambda =\mu^{-1}$ which would be of the order of
macroscopic distances if $m$ is of the order of hadronic masses.
This observation was one of the motivations to propose a non-Newtonian force,
or the fifth force [\cite{ff}].  The constraint $\omega\gsim 500$ would not
apply if $\lambda\ll $ 1 AU.

Another important requirement in the original BD theory is the assumed absence
of the {\em direct} coupling of $\phi$ to the matter fields at the level of the
Lagrangian, though the coupling arises {\em indirectly}.  In fact this
assumption results in the geodesic equation for a matter particle, hence
ensuring WEP.
The purpose of this note is to show that this result is also fragile against
quantum effects.\footnote{Violation of WEP does not necessarily imply the same
of Equivalence Principle (EP) in the more fundamental and theoretical sense;
tangential spacetime at each point of curved spacetime should be Minkowskian.}

For the sake of illustration we consider the matter system of nucleons and the
electromagnetic field.
The analysis is focused upon electromagnetic interaction of protons.  In the
original theory of (\ref{cp_1}), however, the $\phi$-matter coupling occurs
only in the field equation.
This makes it inconvenient to apply the conventional technique of QED.  We then
apply a conformal transformation [\cite{dk}]\footnote{The final results remain
the same in both conformal frames before and after the transformation
in our calculation to the lowest order with respect to $G$, though this is not
the case in general.
The equivalence can be established by introducing $\sigma$ in the
{\em original} conformal frame;
$\phi=\sqrt{\omega/2\pi G}+(1/\sqrt{1+3/2\omega})\sigma$.
Diagonalization corresponding to (\ref{cp_2}) is also useful.  More details
will be discussed elsewhere.}:
\begin{equation}
g_{\mu\nu}\rightarrow g_{*\mu\nu}=\frac{2\pi G}{\omega}\phi^{2}g_{\mu\nu}.
\label{cp_2}\end{equation}
In the new starred conformal frame the Lagrangian (\ref{cp_1}) is cast into the
form
\begin{equation}
{\cal L}=\sqrt{-g_{*}} \left[\;\frac{1}{16\pi G}R_{*}
-\frac{1}{2}g^{\mu\nu}_{*}\partial_{\mu}\sigma\partial_{\nu}\sigma
+L_{*\rm matter}
 \right],  \label{cp_3}\end{equation}
where $\sigma$ is a redefined canonical scalar field related to $\phi$ by
\[
\phi =\sqrt{\frac{\omega}{2\pi G}}e^{\beta\sigma},
\qquad\mbox{with}\qquad\beta=\sqrt{\frac{4\pi G}{3+2\omega}}.
\]

For the matter Lagrangian with the proton field $\psi$ and the electromagnetic
field $A_{\mu}$, the transformed Lagrangian differs in form only in the mass
term:
\begin{equation}
L_{*\rm mass}=-me^{-\beta\sigma}\psibar_{*}\psi_{*},
\label{cp_5}\end{equation}
where $\psi_{*}=(2\pi G/\omega)^{-3/4}\phi^{-3/2}\psi$, while $A_{\mu}$ remains
unchanged.
The exponential factor in (\ref{cp_5}) shows the presence of the direct matter
coupling in the new conformal frame.
In accordance with this, the geodesic equation is modified but in such a
universal manner that WEP remains valid.\footnote{The right-hand side of the
geodesic equation is now
$-\beta(\partial_{\nu}\sigma )(g_{*}^{\mu\nu}+u_{*}^{\mu}u_{*}^{\nu})$,
being independent of properties specific to individual objects.}
In the following we suppress the symbol $*$ to simplify the notation.

We focus upon the linear term
\begin{equation}
L_{\sigma}=\beta m\overline{\psi}\psi\sigma.
\label{cp_6}\end{equation}
This is the coupling to the trace of the matter energy-momentum tensor
$T=-m\overline{\psi}\psi$, with no contribution from $A_{\mu}$.
This interaction, corresponding to a vertex in Feynman diagrams, may be
represented by a ``mass insertion" to a proton line in the limit of a vanishing
momentum transferred to $\sigma$.

We now calculate the one-loop correction to this vertex.  To the lowest-order
approximation with respect to the fine-structure constant $\alpha =e^{2}/4\pi
\approx 1/137$, we consider two types of diagrams: (a) mass insertion in the
{\em internal} proton line of the proton self-energy diagram (se);
(b) mass insertion to the {\em external} proton lines attached to (se).

The contribution from the diagram (se) is represented by
\[
\Sigma(m,i\psh)=A(m)+\left( m+i\psh \right)B(m)
+\left( m+i\psh \right)^{2}\Sigma_{f}(m,i\psh),
\]
where $A=\delta m$ is the self-energy as given by a divergent integral
[\cite{jr}]
\begin{equation}
A=m\frac{\alpha}{2\pi}\left[
-i\frac{1}{\pi^{2}}\int dx(1+x)\int d^{4}k\frac{1}{\left( k^{2}+x^{2}m^{2}
\right)^{2}}
 -\frac{1}{4}\right],
\label{cp_8}\end{equation}
while $B$ represents a wave-function renormalization.

The correction results in replacing $m$ in (\ref{cp_6}) by
${\cal M}$ in which the portion proportional to
$\alpha$ is given by ${\cal M}_{1}={\cal M}_{a}+{\cal M}_{b}$, where
\begin{equation}
{\cal M}_{a}=\left[ \left( m\frac{\partial}{\partial m} \right)
\Sigma(m,i\psh)\right]_{ip\hspace{-0.3em}/ =-m},
\qquad\mbox{and}\qquad {\cal M}_{b}=-mB.
\label{cp_10}\end{equation}
In ${\cal M}_{a}$, $(\partial/\partial m)$ adds another proton line, while $m$
provides the mass at the vertex.
In the calculation of ${\cal M}_{b}$, on the other hand, we have included
$-\delta m$ in the self-energy part.  A cancellation occurs among the terms of
$B$, hence leaving
\begin{equation}
{\cal M}_{1}= m\frac{dA}{dm} = \delta m-\frac{3\alpha}{2\pi}m,
\label{cp_12}\end{equation}
where (\ref{cp_8}) has been used.

The first term $\delta m$ is obviously absorbed into the zeroth order term $m$,
 thus replacing the bare mass by the observed renormalized mass $m +\delta m$.
On the other hand, the second term $-(3\alpha/2\pi)m$ represents an extra
contribution not to be included in the trace.
The fact that this term is not part of mass can be seen by examining $T_{00}$
which is the source of the gravitational field for a {\em static renormalized}
proton:
\begin{eqnarray*}
T_{00}&\approx&\half \psibar\left( \gamma_{0}\overrightarrow{\partial}_{0}
-\overleftarrow{\partial}_{0}\gamma_{0} \right)\psi \nonumber\\
&=&-\half\psibar\left( \overrightarrow{\partial}\hspace{-0.8em}/
-\overleftarrow{\partial}\hspace{-0.8em}/    \right)\psi\nonumber\\
&=&\psibar\left[ \left( m+\delta m\right) -\beta m\sigma+\cdots  \right]\psi,
\label{cp_13}\end{eqnarray*}
where $\cdots$ means other interaction terms.   Notice that due to the Ward
identity, no renormalization is present for $T_{\mu\nu}$ as long as we stay out
 of quantum effects of gravity in 4 dimensions.\footnote{I thank K. Fujikawa
 for the related comments.}  We have retained the $\sigma$-term because it
might behave like a constant in a region where the field $\sigma(\vec{x})$ is
nearly constant.
This term, however, is negligible of the order of $\sqrt{G}$ being entirely
different from the above extra term $-(3\alpha/2\pi)m$.
In this way we find that {\em the extra term gives a difference between what
the tensor gravitational field  feels and what the scalar field feels, hence
the violation of WEP.}
We also observe that the term depends explicitly on $\alpha$, one of the
``internal parameters," beyond the extent to which it is absorbed into the
mass.

We point out that the extra term is {\em finite} in the same context as in the
``trace anomaly".\footnote{Trace anomaly itself is consistent with EP at the
fundamental level.  It can be relevant to the experimental WEP violation only
if $\sigma$ is sufficiently long-ranged.}
WEP which has been designed to be valid classically is induced to be violated
following the manner of ``quantum anomalies."
This finiteness allows us to make a less ambiguous prediction than the
force-range of  $\sigma$.  It is interesting to notice that the finiteness can
also be understood by using an explicit expression
$\delta m=(3\alpha/2\pi)m[ \ln (\Lambda/m)+1/4  ]$, where $\Lambda$ is a
cutoff.  The anomalous term arises simply because a naive dimensional analysis
fails due to the presence of the cutoff.

We now discuss phenomenology.  By assuming a nonzero mass $\mu=\lambda^{-1}$
of $\sigma$, we have a fifth-force-type potential:
\begin{equation}
V_{ij}=-G\frac{M_{i}M_{j}}{r}\left( 1+ \frac{q_{i}q_{j}}{3+2\omega}
{\rm e}^{-r/\lambda}   \right),
\label{cp_14}\end{equation}
as will be read off from (\ref{cp_6}).

According to the present model, the neutral neutron is free from anomaly;
$q_{n}=1$, whereas (\cite{cp_12}) for the proton is translated into
$q_{p}=1-(3\alpha/2\pi).$

Consider a nucleus $i$ with the mass number $A_{i}$ and the atomic number
$Z_{i}$.
Again as an illustration, we choose a simplified model in which a nucleus is
made of nucleons and fields mediating nuclear forces but assumed to {\em
observe} WEP.  We also ignore electrons.
Then $q_{i}$ for the atom/nucleus is given by
\[
q_{i}=1-\frac{3\alpha}{2\pi}\frac{Z_{i}m}{M_{i}}\approx
1-\frac{3\alpha}{2\pi}\frac{Z_{i}}{A_{i}},
\]
where $M_{i}$ is the mass of the nucleus and the proton mass $m$ replaced by
$\approx M_{i}/A_{i}$.

The fractional difference of the acceleration between the two nuclei $i$ and
$j$ toward a source $S$ is then
\[
\frac{\delta a_{ij}}{a_{ij}}=\frac{1}{3+2\omega}
\left( q_{i}-q_{j}\right)q_{S}{\cal F},
\]
where $q_{S}$ of the source may be replaced by the composition-independent
component 1, while ${\cal F}(R,\lambda)$ takes care of the finite force-range;
${\cal F}\rightarrow 1$ for $\lambda \gg R$.

Suppose first that $\lambda\gsim$1AU.  Then the most stringent constraint comes
 from the null experiments by Roll, Krotkov and Dicke, and by Braginski and
Panov [\cite{rkd}].
These result impose the condition $|\delta a_{ij}/a_{ij}|\lsim 10^{-12}$.  This
 translates into
\[
\left|\frac{1}{3+2\omega}\frac{3\alpha}{2\pi}\delta \left( \frac{Z}{A}
 \right)\right|\lsim 10^{-12}.
\]
We find $\delta (Z/A)\approx 0.08$ for Al and Pt/Au, hence giving
$\omega \gsim 1.5\times 10^{8}$, far larger than $\omega \gsim 500$, as
obtained from the solar-system experiments.

If $\lambda\ll$1AU, the free-fall experiments on the Earth [\cite{km}] then
give the lower bounds of $\omega$;
460, $5.2\times 10^{3}$ and $2.5\times 10^{4}$ for the assumed values of
$\lambda$; 100km, $10^{3}$km and $10^{4}$km, respectively.
They still tend to be more stringent than 500.

If $\lambda$ is as small as 1m or even less, we resort to the
composition-dependent experiments using terrestrial sources.
Some of the recent measurements [\cite{ws}] reached the accuracy
$\sim 10^{-11}$ for the upper bound of the fractional acceleration difference
between Cu and Be, giving $\omega \gsim 2.2\times 10^{6}$.
This is true for $\lambda\gsim 1$ m, thus surpassing the limits set by the
free-fall experiments.

The amount of composition-dependence as represented by $3\alpha/2\pi$ is quite
large as compared with what has been expected from the baryon-number coupling.
Combining this with extremely accurate tests of WEP is the reason why we
``improved" the bound of $\omega$ by many orders of magnitude.
The new constraint will affect many analyses attempted so far on the BD theory.
  On the other hand, very large $\omega$ may be avoided if $\lambda$ is
  sufficiently small, perhaps shorter than 1m.

Our analysis was based on a simplified model in which composition-dependence
comes only from the anomaly in protons.
For this reason the results are tentative, mainly to demonstrate the importance
 of the issue.
Obviously other fields should be taken into account.  Even the electromagnetic
interaction of the neutron through the anomalous magnetic moment may not be
ignored.
It might be better to apply the same technique to the quarks with the gauge
fields at a more fundamental level.
With such elaborations, however, the effect of roughly the same size still
seems unavoidable unless some cancellation mechanism is discovered to work.

A possible way to avoid unnaturally large values of $\omega$ is to invoke a
suppression mechanism as advocated by Damour and Nordtvedt [\cite{dn}].
Even with the suggested modification of the nonminimal coupling, however, the
anomalous effect should be still present to be included in the analysis.
In this connection we also point out that the argument depends crucially on the
 assumption of a {\em standard} mass term in the original conformal frame in
which the Lagrangian is given by (\ref{cp_1}).
Unification models suggest the $\phi$-dependence of the mass terms.  An example
 was studied in which the ``mass" tends asymptotically to a constant value as
 the Universe expands [\cite{fn}].\footnote{According to this model, the
strength of the $\sigma$ coupling is suppressed by the age of the Universe.}
Further study of various versions of the scalar-tensor theory in conjunction
with the experimental efforts will benefit the development toward unification.

Finally we add that the mass insertion to the photon self-energy part yields
the anomalous coupling
\[
L_{\sigma\gamma}=-\frac{2\alpha}{3\pi}\beta\frac{1}{4}F_{\mu\nu}F^{\mu\nu}
\sigma,
\]
which is also finite.  This is interesting because it is not a trace coupling,
certainly violating WEP.
Unfortunately, the effect through the nuclear Coulomb energy is numerically
negligible compared with that of the proton anomaly.

I thank Y.M. Cho for discussions at the early stage of the work.  I also
acknowledge useful discussions with K. Kuroda and R. Newman.

\bigskip
\begin{center}
{\Large\bf References}
\end{center}
\setlength{\itemsep}{0.2cm}
\begin{enumerate}
\item\label{jbd}C. Brans and R.H. Dicke, Phys. Rev. {\bf 124}, 925 (1961).
\item\label{ff}Y. Fujii, Nature Phys. Sci, {\bf 234}, 5\ (1971); Phys. Rev.
{\bf D9}, 874 (1974).
See, for the latest review, E. Fischbach and C. Talmadge, Nature, {\bf 356},
207\ (1992).
\item\label{dk}R.H. Dicke, Phys. Rev. {\bf 125}, 2163\ (1962).
\item\label{jr}See, for example, J.M. Jauch and F. Rohrlich, {\sl The Theory of
 Photons and Electrons,} Addison-Wesley, 1954: C. Itzykson and J.-B. Zuber,
{\sl Quantum Field Theory,} McGraw-Hill, 1985.
\item\label{rkd}P.G. Roll, R. Krotkov and R.H. Dicke, Ann. Phys. (N.Y.)
{\bf 26}, (1964)\ 442: V.B. Braginski and V.I. Panov, Sov. Phys. JETP {\bf 34},
 (1972)\ 463.
\item\label{km}T.M. Niebauer, M.P. McHugh and J.E. Faller, Phys. Rev. Lett.
{\bf 59}, 609 (1987): K. Kuroda and N. Mio, Phys. Rev. Lett. {\bf 62}, 1941
(1989).    The results in the text were obtained from the second reference.
\item\label{ws}E.G. Adelberger et al., Phys. Rev. {\bf D42}, 3267 (1990).
\item\label{dn}T. Damour and K. Nordtvedt, Phys. Rev. Lett. {\bf 70}, 2217
(1993):
See also, T. Damour and A.M. Polyakov, The String Dilaton and a Least Coupling
Principle, preprint.
\item\label{fn}Y. Fujii and T. Nishioka, Phys. Rev. {\bf D42}, 361 (1990).
\end{enumerate}

\end{document}